\documentclass[doublecol]{epl2}

\pdfoutput=1

\usepackage{graphicx}
\usepackage{subfigure}
\usepackage{hyperref}
\hypersetup{
  colorlinks,
  citecolor=red,
  filecolor=black,
  linkcolor=blue,
  urlcolor=black
}
\usepackage{amsmath}
\usepackage{amsthm}
\usepackage{amssymb}

\allowdisplaybreaks[1]

\def\SZ{S_0}
\def\deltad{\delta_\mathrm{D}}
\def\taur{\tau_\mathrm{R}}
\def\taub{\tau_\mathrm{B}}

\newcommand{\dd}[1]{\!\mathrm{d}#1\,}
\newcommand{\td}[1]{\frac{\mathrm{d}}{\mathrm{d}t}}

\newcommand{\floor}[1]{{\lfloor\hspace{-.25mm} #1 \hspace{-.25mm}
    \rfloor}}

\newcommand{\fref}[1]{fig.~\ref{#1}}

\title
{Transport properties of L\'evy walks: an analysis in terms of
  multistate processes}
\shorttitle{Transport properties of L\'evy walks in terms of
  multistate processes}

\author{
  Giampaolo Cristadoro%
  \inst{1} 
  \and 
  Thomas Gilbert%
  \thanks{E-mail: \email{thomas.gilbert@ulb.ac.be}}
  \inst{2} 
  \and
  Marco Lenci
  \inst{1,3}
  \and 
  David P.~Sanders
  \inst{4}}  
\shortauthor{G. Cristadoro \etal}

\institute{                    
  \inst{1}Dipartimento di Matematica, Universit\`a di Bologna, 
  Piazza di Porta S. Donato 5, 40126 Bologna, Italy.\\
  \inst{2}Center for Nonlinear Phenomena and Complex Systems,
  Universit\'e Libre  de Bruxelles, C.~P.~231, Campus Plaine, B-1050
  Brussels, Belgium.\\
  \inst{3}Istituto Nazionale di Fisica Nucleare, Sezione di
  Bologna, Via Irnerio 46, 40126 Bologna, Italy.\\ 
  \inst{4}Departamento de F\'isica, Facultad de Ciencias, Universidad
  Nacional Aut\'onoma de M\'exico,  Ciudad Universitaria,
  04510 M\'exico D.F., Mexico.
}

\abstract{
  Continuous time random walks combining diffusive and ballistic
  regimes are introduced to describe a class of L\'evy walks on 
  lattices. By including exponentially-distributed waiting times
  separating the successive jump events of a walker, we are led to
  a description of such L\'evy walks in terms of multistate processes
  whose time-evolution is shown to obey a set of coupled delay differential
  equations. 
  Using simple arguments, we obtain asymptotic solutions to
  these equations and rederive the scaling laws for the mean
  squared displacement of such processes. 
  Our calculation includes the computation of all relevant transport
  coefficients in terms of the parameters of the models.
}

\begin{document}

\maketitle

Random walks described by L\'evy flights give rise to complex
diffusive processes \cite{Haus:1987p513, Weiss:1994RandomWalk,
  Krapivsky:2010ti, klafter:2011first} and have found many
applications in physics and beyond \cite{Shlesinger:1995uk,
  Klages:2008p11988, Denisov:2012PhRvE85, Mendez:2013stochastic}.
Whereas the random walks associated with Brownian motion are
characterized by Gaussian propagators whose variance grows linearly in
time, the propagators of L\'evy flights have infinite variance
\cite{Shlesinger:1993fw, Klafter:1996PhysTod, Metzler:2000PhysRep};  
they occur in models of random walks such that the
probability of a long jump decays slowly with its length
\cite{Weiss:1983RandomWalks}. 

By  considering the propagation time between
the two ends of a jump, one 
obtains a class of models known as L\'evy walks
\cite{Geisel:1985p8023, Shlesinger:1985p18111, Shlesinger:1987p12279,
  Klafter:1987Stochastic, Blumen:1989PRA3964, Zumofen:1993p804,
  shlesinger:1999above}. A  L\'evy 
walker thus follows a continuous path between the 
two end points of every jump, performing each in a
finite time; instead of having an infinite mean squared displacement, as
happens in a L\'evy flight whose jumps take place instantaneously, a
L\'evy walker moves with finite velocity and, ipso facto, has a finite
mean squared displacement, although it may increase faster
than linearly in time.

A L\'evy flight is characterized by its probability density of jump
lengths $x$,
or \emph{free paths}, which we denote $\phi(x)$. It is assumed to have
the asymptotic scaling, $\phi(x) \sim x^{-\alpha - 1}$, whose exponent,
$\alpha > 0$, determines whether the moments of the
displacement are finite. In particular, for $\alpha \leq 2$, the variance
diverges.  

In the framework of continuous time random walks
\cite[Chs. 10 \& 13]{Shlesinger:1995uk}, a
probability distribution $\Phi(\vect{r}, t)$ of making a
displacement $\vect{r}$ in a time $t$ is introduced, such that, for
instance, in the so-called velocity picture, 
$\Phi(\vect{r}, t) = \phi(|\vect{r}|) \deltad(t - |\vect{r}|/v)$, 
where $v$ denotes the constant speed of the particle and  $\deltad(.)$
is the Dirac delta function. Considering the Fourier-Laplace
transform of the propagator of this process, one obtains, in terms of
the parameter $\alpha$, the following scaling laws for the
mean squared displacement after time $t$ \cite{Geisel:1985p8023, 
  Wang:1992PhysRevA.45.8407},  
\begin{equation}
  \langle r^2 \rangle_t \sim
  \begin{cases}
    t^2\,, & 0 < \alpha < 1\,, \\
    t^2/\log t\,, & \alpha = 1\,,\\
    t^{3 - \alpha}\,, & 1 < \alpha < 2\,,\\
    t \log t\,, & \alpha = 2\,,\\
    t\,, & \alpha > 2\,.
  \end{cases}
  \label{eq:scalings}
\end{equation}

In this Letter, we consider L\'evy walks on lattices and generalize the
above description, according to which a new jump event takes place as
soon as the previous one is completed, to include an
exponentially-distributed waiting 
time which separates successive jumps. This induces a
distinction between the states of particles which are in the process
of completing a jump and those that are waiting to start a new
one. 
As shown below, such considerations lead to a theoretical
formulation of the model as a multistate generalized master equation
\cite{Montroll:1965p243, Kenkre:1973em, Landman:1977PNAS430}, which
translates into a set of coupled delay differential equations for the
corresponding 
distributions. 

The physical motivation for the inclusion of an
exponentially-distributed waiting time between successive jump 
events stems, for instance, in the framework of chaotic scattering,
from the time required to escape a fractal repeller
\cite{Ott:1993p11313, Gaspard:1998book}, or, more generally, the
time spent in a chaotic transient \cite{Kantz:1985p4257}. In the
framework of active transport, such as dealing with the motion 
of particles embedded within living cells \cite{Gal:2010PRE81}, such
waiting times may help model the complex process related to changes
in the direction of propagation of such particles. This is also
relevant to laser cooling experiments \cite{Bardou:2001CUP}, where a
competition in the damping and increase of atomic momenta induces a
form of random walk in momentum space. The times spent by atoms in
small momenta states typically follow exponential distributions.

The stop-and-go patterns of random walkers thus generated have been
studied in the context of animal foraging \cite{Obrien:1990search}. 
Such search strategies have been termed saltatory. In contrast to
classical strategies, according to which 
animals either move while foraging or stop to ambush their
prey, a saltatory searcher alternates between scanning
phases, which are performed diffusively on a local scale, and
relocation phases, during which motion takes place without search.
Examples of such intermittent behaviour have been identified in a
variety of animal species \cite{Kramer2001:behavioral,
  Mashanova:2010evidence},  as well as in intracellular processes such
as proteins binding to DNA strands \cite{Halford:2004site}. Visual
searching patterns whereby information is extracted through a cycle of
brief fixations interspersed with gaze shifts\cite{ludwig:2014foveal}
provide another illustration in the context of neuroscience. One can
also think of applications to sociological processes, for instance
when interactions between individuals is sampled at random times,
independent of the underlying process \cite{Brockmann:2006p8047}. 

From a mathematical perspective, an important question that arises in
the framework of foraging is that of optimal strategies
\cite{Viswanathan:2011physics}. As reported in
\cite{Viswanathan:1999optimizing}, L\'evy flight motion can, under
some conditions on the nature and distribution of targets, emerge as
an optimal strategy for non-destructive search, i.e., when targets can
be visited infinitely often. For destructive searches on the other
hand, intermittent search strategies with exponentially-distributed
waiting times provide an alternative to L\'evy search strategies,
which turns out to minimize the search time
\cite{Benichou:2011intermittent}. The processes we 
analyze in this Letter, although they are restricted to motion on
lattices, can be thought of as extensions of intermittent search
processes to power-law distributed relocation phases which are typical
of L\'evy search strategies, thus opening a new perspective.

We show below that, inasmuch as the 
dispersive properties are
concerned,  a complete characterisation of the process can be
obtained, which reproduces the scaling laws \eqref{eq:scalings}, as
well as yields the corresponding transport coefficients, whether
normal or anomalous.
These results also elucidate the incidence of exponential waiting
times on these coefficients. 

\section{L\'evy walks as multistate processes}

We call \emph{propagating} the state of a particle which is in the
process of completing a jump. In contrast, the state of a particle
waiting to start a new jump is called \emph{scattering}
\footnote{B\'enichou et \emph{al.} \cite{Benichou:2011intermittent}
  refer to these two states as respectively ballistic and diffusive.}.  
Whereas particles switch from propagating to scattering states as they
complete a jump, particles in a scattering state can make
transitions to  both scattering and propagating states; as soon as their 
waiting time has elapsed, they move on to a neighbouring site and,
doing so, may switch to a propagating state and carry their motion on to
the next site, or start anew in a scattering state. 

We consider a $d$-dimensional cubic lattice of individual cells
$\vect{n}\in\mathbb{Z}^d$. The state of a walker at position
$\vect{n}$ and time $t$ can take on a countable number of different
values, specified by two integers, $k\geq 0$ and $j \in
\{1,\dots,z\}$, where $z \equiv 2d$ is the coordination number of the
lattice. Scattering states are labeled by the state $k=0$ and propagating
states by the pair $(k,j)$, such that $k\geq 1$ counts the remaining
number of lattice sites the particle has to travel in direction $j$
to complete its jump. 

Time-evolution proceeds as follows. After a
random waiting time $t$, exponentially-distributed with mean 
$\taur$, a particle in the scattering state $k=0$ changes its state to
$(k, j)$ with probability $\rho_{k}/z$, moving its location from site
$\vect{n}$ to site $\vect{n} + \vect{e}_j$. Conversely,
particles which are at site $\vect{n}$ in a propagating state $(k,
j)$, $k\geq1$, jump to site $\vect{n} + \vect{e}_j$, in time
$\taub$, changing their state to
$(k-1, j)$. 

The waiting time density of the process is the 
function
\begin{equation}
  \psi_k(t) = 
  \begin{cases}
    \taur^{-1} e^{- t/\taur}\,, & k =0\,,\\
    \deltad(t -\taub)\,, & k \neq 0\,.
  \end{cases}
  \label{eq:psik}
\end{equation}
When a step takes place, the
transition probability to go from state $(k,j)$ to state $(k',j')$ is
\begin{equation}
  \mathsf{p}_{(k,j),(k',j')} = 
  \begin{cases}
    \rho_{k'}/z\,,&  k =0\,,\\
    \delta_{k-1,k'} \delta_{j,j'} \,, & k \neq 0\,,
  \end{cases}
  \label{eq:Pkj}
\end{equation}
where $\delta_{.,.}$ is the Kronecker symbol.

For definiteness, we consider below the following simple
parameterisation of the transition probabilities,
\begin{equation}
  \rho_k =
  \begin{cases}
    1 - \epsilon\,,  & k = 0\,,\\
    \epsilon
    \left[k^{-\alpha}- (k+1)^{-\alpha}\right]\,,
    & k\geq1\,,
  \end{cases}
  \label{eq:defrhok}
\end{equation}
in terms of the parameters $0< \epsilon < 1$, which weights scattering
states relative to propagating ones, and $\alpha>0$, the asymptotic
scaling parameter of free path lengths. 

\section{Master equation}

The probability distribution of particles at site $\vect{n}$ and
time $t$, $P(\vect{n}, t)$, is a sum of the distributions over the
scattering states, $P_0(\vect{n}, t)$, and propagating states,
$P_{k,j}(\vect{n}, t)$, $k\geq 1$ and $1 \leq j \leq z$. According to 
eqs.~\eqref{eq:psik} and \eqref{eq:Pkj}, changes in the distribution of
$(k,j)$-states, $k\geq 1$, in cell $\vect{n}$ arise from particles
located at cell $\vect{n} - \vect{e}_j$ which make
a transition from either state $0$ or state $(k+1, j)$. Since the
latter transitions can be traced back to changes in the distribution
of $(k+1, j)$-states in cell $\vect{n} - \vect{e}_j$ 
at time $\taub$ earlier, we can write\footnote{The possible addition
  of source terms into this expression will not be considered here.} 
\begin{align}
  \partial_t P_{k,j} (\vect{n}, t)  &-   
  \partial_t P_{k+1,   j}(\vect{n} - \vect{e}_j, t - \taub)   
  \cr
  =&
  \frac{\rho_{k}}{z \taur} [ P_{0}(\vect{n} - \vect{e}_j, t)
  - P_{0}(\vect{n} - \vect{e}_j, t - \taub)],
\end{align}
which accounts for the fact that a positive $0$-state contribution at
time $t$ becomes a negative one at time $t + \taub$.
Applying this relation recursively, we have
\begin{widetext}
  \begin{equation}
    \partial_t P_{k,j}(\vect{n}, t)  =  
    \frac{1}{z \taur}\sum_{k' = 1}^\infty   \rho_{k + k'-1} 
    \Big[
    P_{0}(\vect{n} - k' \vect{e}_{j}, t - (k'-1) \taub)
    - P_{0}(\vect{n} - k' \vect{e}_{j}, t - k' \taub)
    \Big]\,.
    \label{eq:dtPkj}
  \end{equation}
\end{widetext}
\begin{floatequation}
  \mbox{\textit{See eq.~\eqref{eq:dtPkj} next page.}}
\end{floatequation}
\setcounter{equation}{6}
Terms lost by $(1,j)$-states in cells $\vect{n} - \vect{e}_j$, $j
= 1,\dots, z$, are gained by the $0$-state in cell $\vect{n}$, which also
gains contributions from $0$-state transitions. Since the scattering state
also loses particles at exponential rate $1/\taur$, we have
\begin{align}
  \!\!  \!
  \partial_t P_0(\vect{n}, t)  & =  
  \frac{1}{z \taur} \sum_{j = 1}^{z}
  \sum_{k=0}^\infty   \rho_{k} 
  P_{0}(\vect{n} - (k+1) \vect{e}_j, t - k \taub)
  \cr
  & \quad  
  - \frac{1}{\taur}P_{0}(\vect{n}, t)
  \,.
  \label{eq:dtP0}
\end{align}
It is straightforward to check that eqs.~\eqref{eq:dtPkj} and
\eqref{eq:dtP0} are consistent with conservation of
probability\footnote{A simplification occurs if one considers the
  distribution of propagating states in direction $j$,
  $P_{j}(\vect{n}, t) = \sum_{k=1}^{\infty} P_{k,j}(\vect{n},
  t)$. Using eq.~\eqref{eq:defrhok},  the time-evolution of this
  quantity is  $\partial_t P_{j}(\vect{n}, t)  =   \epsilon/(z \taur)
  \sum_{k = 1}^\infty   k^{-\alpha} [
  P_{0}(\vect{n} - k \vect{e}_{j}, t - (k-1) \taub)
  - P_{0}(\vect{n} - k \vect{e}_{j}, t - k \taub)
  ]$.}, $\sum_{\vect{n}} P(\vect{n}, t) = 1$. 

\section{Fraction of scattering particles}

As discussed below, an important role is played by the overall
fraction of particles in the scattering state, 
$\SZ(t) \equiv \sum_{\vect{n}\in\mathbb{Z}^d} P_{0}(\vect{n}, t)$.
From eq.~\eqref{eq:dtP0}, this quantity is found to obey the following
linear delay differential equation, 
\begin{equation}
  \taur \dot \SZ(t) 
  = \sum_{k=1}^\infty \rho_k \SZ(t - k \taub) -
  \epsilon \SZ(t) \,.
  \label{eq:dtS0gen}
\end{equation}
Given initial conditions, e.g. $\SZ(t) = 0$, $t<0$, and $\SZ(0) = 1$
(all particles start in a scattering state), this equation can be
solved by the method of steps
\cite{Driver:1977ordinary}.
Because the sum of the coefficients on the right-hand side of
eq.~\eqref{eq:dtS0gen} is zero, the solutions are asymptotically
constant and can be classified in terms of the parameter $\alpha$.   

\begin{subequations}
  \label{eq:S0largetime}
  For $\alpha>1$, the average return time to the  $0$-state,
  $\sum_{k=0}^\infty \rho_k (\taur + k \taub)$, is finite and given in
  terms of the Riemann zeta function,  since $\sum_{k=0}^\infty
  k \rho_k = \epsilon  \zeta(\alpha)$. The process is thus
  positive-recurrent and we have  
  \begin{equation}
    \lim_{t\to\infty} \SZ(t) = \frac{\taur}
    {\taur + \epsilon \taub \zeta(\alpha)}
    \qquad (\alpha >1)\,.
    \label{eq:S0largetime_agt1}
  \end{equation}
  In the remaining range of  parameter values, $0< \alpha \leq 1$, the
  process is null-recurrent: the average return time to the $0$-state
  diverges and $ \lim_{t\to\infty} \SZ(t) = 0$. If $\alpha \neq 1$, the decay is 
  algebraic, 
  \begin{equation}
    \lim_{t\to\infty}  (t/\taub)^{1 - \alpha} \SZ(t) 
    = \frac{\sin (\pi\alpha)} 
    {\pi \epsilon} \frac{\taur}{\taub} \qquad (0<\alpha <1)\,,
    \label{eq:S0largetime_alt1}
  \end{equation}
  which can be obtained from a result due to Dynkin 
  \cite{Dynkin:1961IMS171};
  see also Refs.~\cite[Vol.~2, \S~XIV.3]{feller1971introduction} and
  \cite[\S~4.4]{Bardou:2001CUP}.
  The case $\alpha = 1$ is a singular limit with logarithmic decay,   
  \begin{equation}
    \lim_{t\to\infty} \log(t/\taub) \SZ(t)  = \frac{1}{\epsilon} 
    \frac{\taur}{\taub} \qquad
    (\alpha = 1)\,.
    \label{eq:S0largetime_aeq1}
  \end{equation}
\end{subequations}

\section{Mean squared displacement}

Assuming an initial position at the origin, the second moment of the
displacement is $\langle n^2 \rangle_t =
\sum_{\vect{n}\in\mathbb{Z}^d} n^2 P(\vect{n},t)$. Its
time-evolution is obtained by differentiating this expression with
respect to time and substituting eqs.~\eqref{eq:dtPkj} and
\eqref{eq:dtP0}, 
\begin{align}
  \taur \td{t} \langle n^2 \rangle_t 
  & = \SZ(t) + \epsilon
  \sum_{k=1}^\infty \frac{2k+1} {k^{\alpha}} \SZ(t- k \taub)\,,
  \label{eq:ddtsumn2}
\end{align}
where, using eq.~\eqref{eq:defrhok}, we made use of the identity $\sum_{j
  = k}^\infty \rho_{j} = 1$ for $k=0$ and $\epsilon k ^{-\alpha}$
otherwise. 
The time-evolution of the second moment is thus obtained
by integrating the fraction of $0$-state particles,
\begin{equation}
  \taur \langle n^2 \rangle_t 
  =  \int_0^{t}\dd{s} \SZ(s) + 
  \epsilon \sum_{k = 1}^{\floor{t/\taub}} 
  \frac{2k+1}{k^{\alpha}} 
  \int_0^{t-k \taub}\dd{s} \SZ(s)\,,
  \label{eq:sumn2}
\end{equation}
where, assuming the process starts at $t=0$, we set $\SZ(t) = 0$ for
$t < 0$. 

\begin{figure}[htb]
  \centering
  \subfigure[~Normal diffusion, $\alpha = 5/2$]{
    \includegraphics[width=.4\textwidth]
    {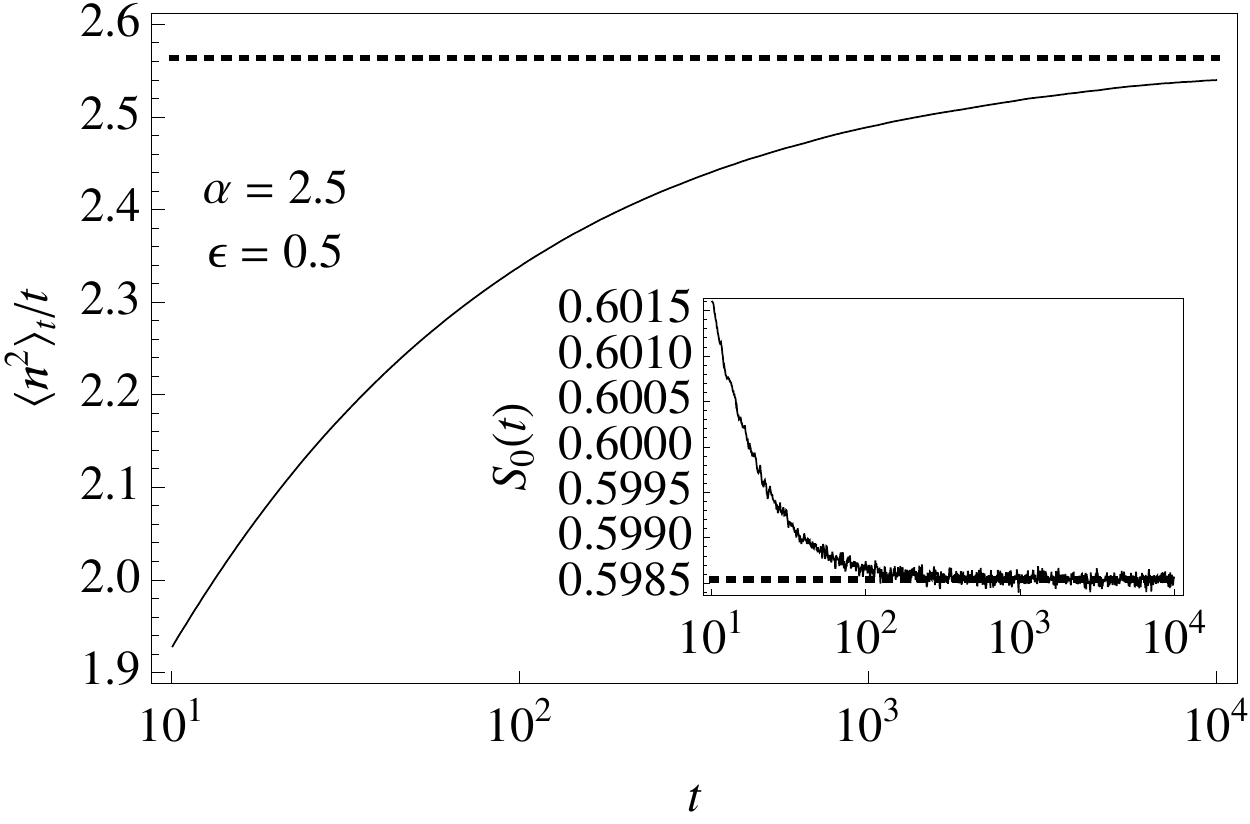}
  }
  \subfigure[~Weak super-diffusion, $\alpha = 2$]{
    \includegraphics[width=.4\textwidth]
    {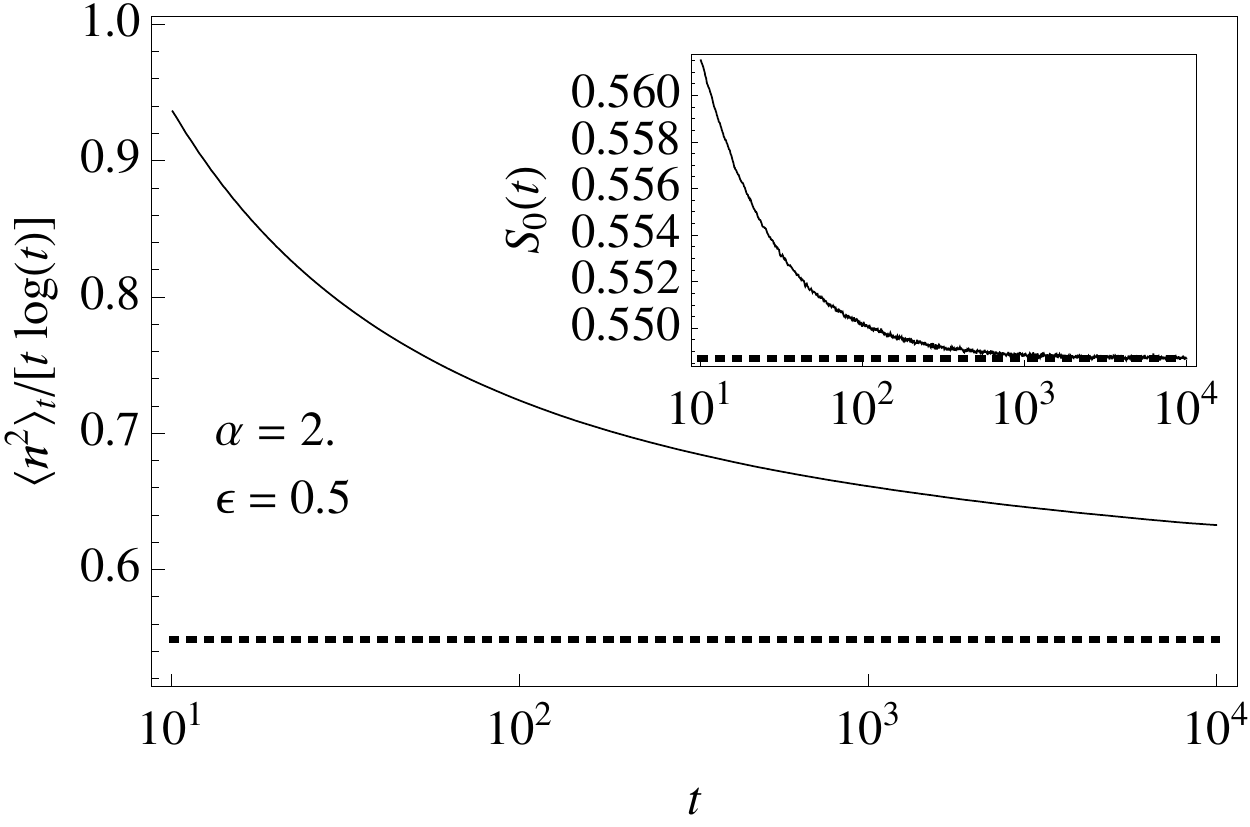}
  }
  \subfigure[~Super diffusion, $\alpha = 3/2$]{
    \includegraphics[width=.4\textwidth]
    {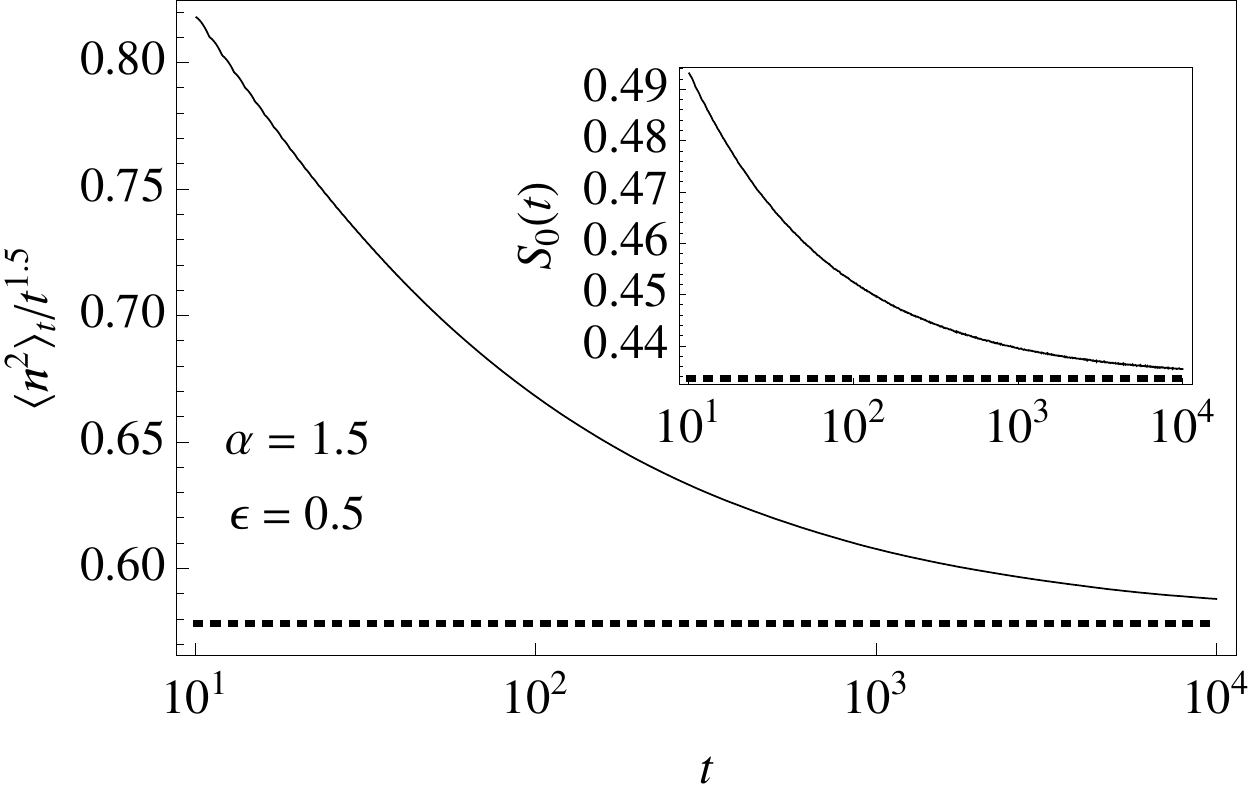}
  }
  \caption{Examples of numerical computations of 
    $\langle n^2\rangle_t$ for parameters values 
    $\alpha >1$, rescaled by their respective asymptotic scalings with
    respect to time ($\epsilon = 1/2$ in all cases).  The dotted lines
    correspond to  eq.~\eqref{eq:msd_posrecurrent}. The insets
    show the evolution of the fraction of scattering states towards their
    asymptotic values, given by eq.~\eqref{eq:S0largetime_agt1}.} 
  \label{fig:msd_posrecurrent}
\end{figure}

\begin{figure}[htb]
  \centering
  \subfigure[~Ballistic diffusion, $\alpha = 1/2$]{
    \includegraphics[width=.4\textwidth]
    {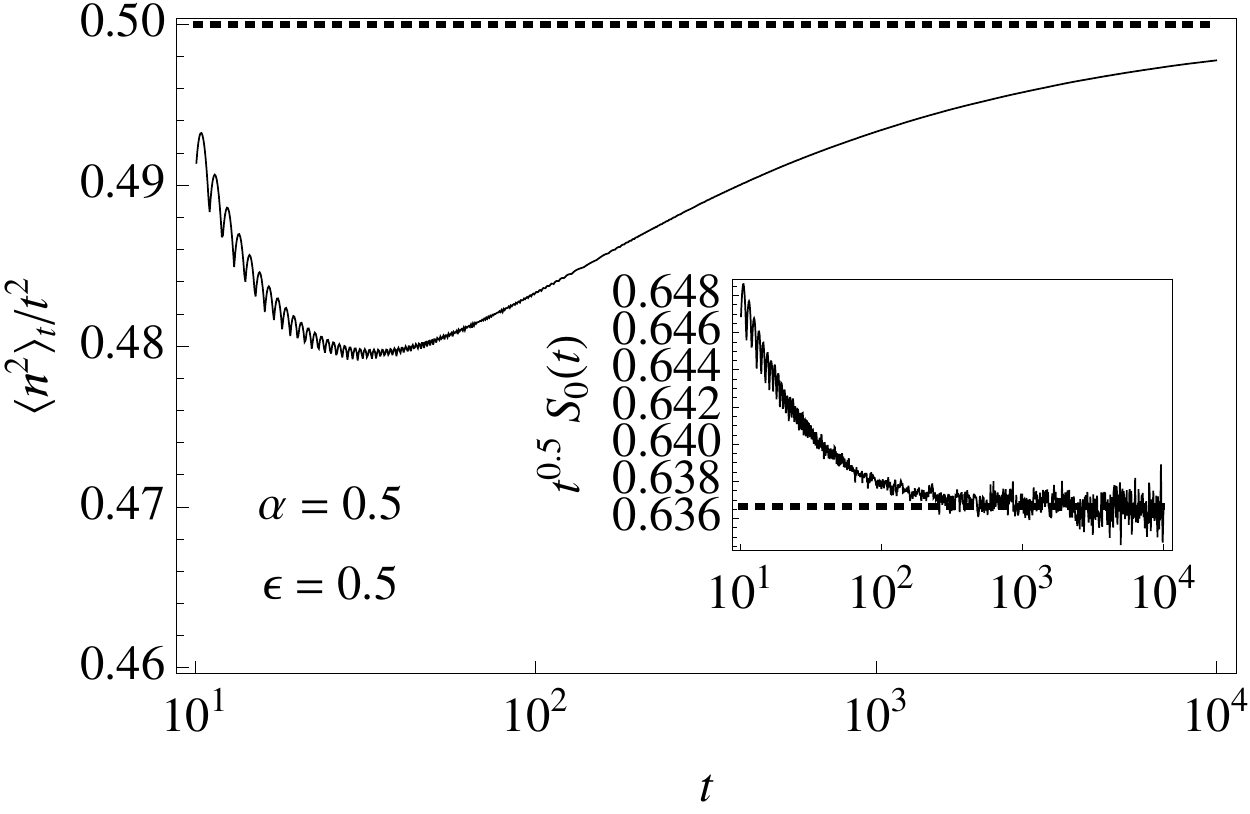}
    \label{fig:msdaneg}
  }
  \subfigure[~Sub-ballistic diffusion, $\alpha = 1$]{
    \includegraphics[width=.4\textwidth]
    {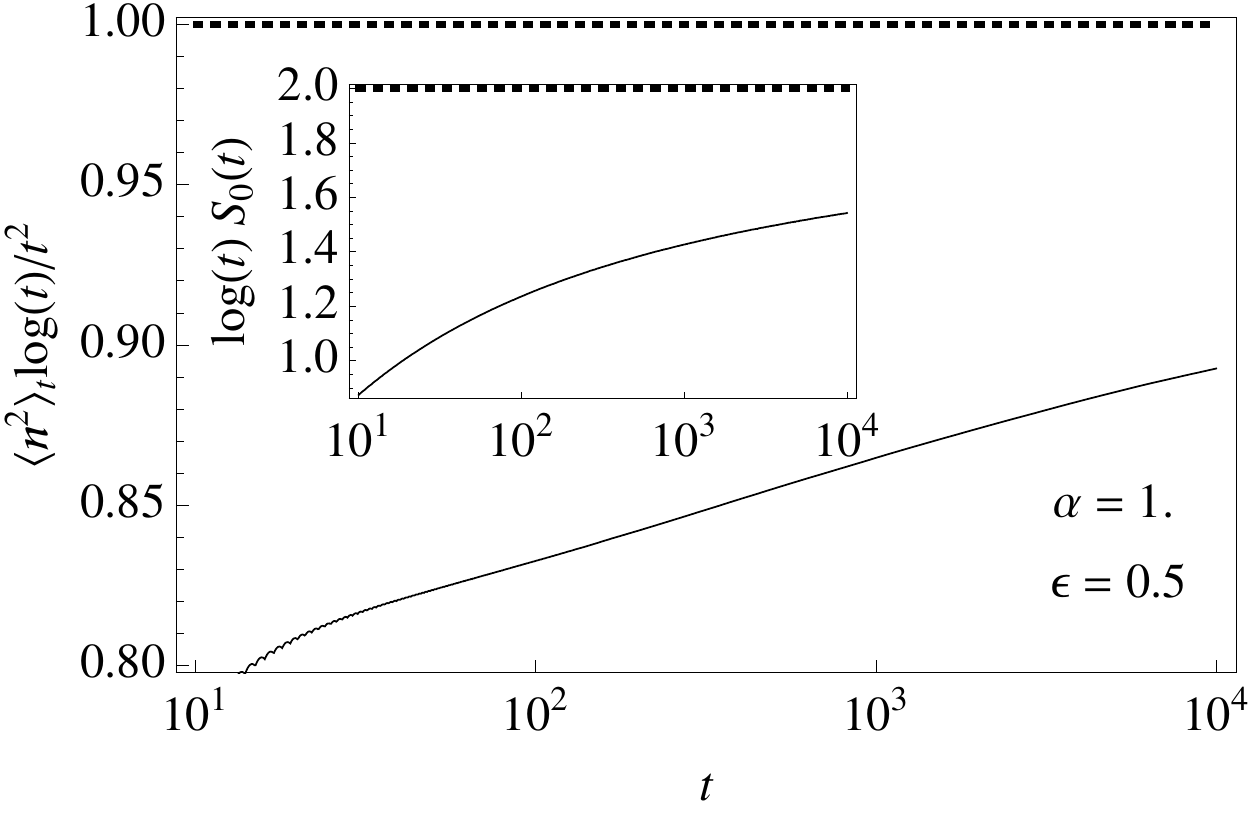}
    \label{fig:msdazero}
  }
  \caption{Same as \fref{fig:msd_posrecurrent} for the range of parameters
    $0 < \alpha \leq 1$, with comparisons to 
    eq.~\eqref{eq:msd_nullrecurrent} and, in the insets,
    eqs.~\eqref{eq:S0largetime_alt1} and \eqref{eq:S0largetime_aeq1}. 
  }
  \label{fig:msd_nullrecurrent}
\end{figure}

As emphasized earlier, equation \eqref{eq:dtS0gen} can be solved
analytically given initial conditions on the state of walkers. By
extension, so can equation \eqref{eq:sumn2}, thus providing an exact
time-dependent expression of the mean squared displacement. This is
particularly useful when one wishes to study transient regimes and the
possibility of a crossover between different scaling behaviours, or
indeed when the asymptotic regime remains experimentally or
numerically unaccessible. The analytic expression of the mean squared
displacement  and the issue of the transients will be studied
elsewhere. Here, we focus on the asymptotic regime, i.e., $t\gg \taub$. 

Substituting the asymptotic expressions 
\eqref{eq:S0largetime}, into
eq.~\eqref{eq:sumn2}, we retrieve the regimes described by
eq.~\eqref{eq:scalings} and obtain the corresponding coefficients.

\begin{subequations}
  \label{eq:msd}
  Starting with the positive-recurrent regime, $\alpha > 1$,
  eq.~\eqref{eq:S0largetime_agt1}, we have the three asymptotic regimes,
  $t\gg \taub$, 
  \begin{align}
    \langle n^2 \rangle_t
    & \simeq 
    \frac{t}{\taur + \epsilon \taub \zeta(\alpha)}
    \cr
    &
    \quad
    \times
    \begin{cases}
      1 + \epsilon[ \zeta(\alpha) + 2  \zeta(\alpha - 1) ]\,, &
      \alpha > 2\,,
      \\
      2 \epsilon \log (t/\taub) \,, &
      \alpha = 2\,,
      \\
      \frac{2 \epsilon}{(2 - \alpha) (3 - \alpha)}
      (t/\taub)^{2 - \alpha}\,, &
      1 < \alpha < 2\,.
    \end{cases}
    \label{eq:msd_posrecurrent}
  \end{align}
  Whereas the first regime, $\alpha > 2$, yields normal diffusion, the
  other two correspond, for $\alpha = 2$, to a weak form of
  super-diffusion, and, for $1 < \alpha < 2$,  to super-diffusion, such that
  the mean squared displacement grows with a power of time $3 
  - \alpha > 1$, faster than linear\footnote{Equation
    \eqref{eq:msd_posrecurrent} assumes $\epsilon > 0$. If 
    one takes the limit $\epsilon\to 0$, sub-leading terms may become
    relevant. In particular, when $\epsilon = 0$, normal diffusion is
    recovered and the right-hand side of \eqref{eq:msd_posrecurrent} is
    $t/\taur$ for all $\alpha$.}.   

  Ballistic diffusion occurs in the null-recurrent regime of the parameter,
  $0 < \alpha \leq 1$. Using eqs.~\eqref{eq:S0largetime_alt1} and
  \eqref{eq:S0largetime_aeq1}, we find  
  \begin{equation}
    \langle n^2 \rangle_t 
    \simeq 
    \frac{t^2}{\taub^2}
    \begin{cases}
      1/\log (t/\taub)\,,
      & \alpha = 1\,,\\
      1 - \alpha\,, & 0 < \alpha < 1\,.
    \end{cases}
    \label{eq:msd_nullrecurrent}
  \end{equation}
\end{subequations}

The asymptotic regimes described by eqs. (12) generalize to
continuous-time processes similar results found in the context of
countable Markov chains applied to discrete time processes
\cite{Wang:1993p11302}. They can also be 
compared to results obtained in Ref.~\cite{Zumofen:1993PowerSpectra}.
Although the L\'evy walks considered by these authors do not include
exponentially-distributed waiting times separating successive
propagating phases, our results are rather similar to theirs; the only actual
differences arise in the regime of normal diffusion, $\alpha > 2$.

In figs.~\ref{fig:msd_posrecurrent} and \ref{fig:msd_nullrecurrent},
the asymptotic results \eqref{eq:msd} are compared
to numerical measurements of the mean squared  displacement of the
process defined by eqs.~\eqref{eq:psik}, \eqref{eq:Pkj} and the
transition probabilities  \eqref{eq:defrhok}. Timescales were set to
$\taur \equiv \taub \equiv 1$ and the lattice dimension to $d=1$. The
algorithm is based on a  classic  
kinetic Monte  Carlo algorithm \cite{Gillespie:1976p296}, which
incorporates the  possibility of a  ballistic propagation of particles
after they undergo a transition from a scattering to a propagating state.
For each realisation, the initial state is taken to be
scattering. Positions are measured at regular intervals on a
logarithmic time scale for times up to $t = 10^4 \taur$. Averages are
performed over sets of $10^8$ trajectories. 

\section{Concluding remarks}

The specificity of our approach to L\'evy walks lies in the inclusion
of exponentially-distributed waiting times that separate
successive jumps. This additional feature induces a natural
description of the process in terms of multiple propagating and scattering
states whose distributions evolve according to a set of coupled delay 
differential equations. 

The mean squared displacement of the process depends on the
distribution of free paths and boils down to a simple
expression involving time-integrals of the fraction of
scattering states. Using straightforward arguments, precise asymptotic
expressions were obtained for this quantity, which reproduce the
expected scaling regimes \cite{Geisel:1985p8023,
  Wang:1992PhysRevA.45.8407}, and provide values 
of the diffusion coefficients, whether normal or anomalous. 

Our results confirm that, in the null-recurrent regime of ballistic
transport, scattering events, are unimportant. Furthermore, these
events do not modify the exponent of the mean squared displacement in
the positive recurrent regime; in other words, the addition of a
scattering phase has no incidence on the scaling exponents.  
In this regime, however, the transport coefficients, whether normal
or anomalous, depend on the details of the model, underlying the
relevance of pausing times that separate long flight events, for example,
in the context of animal foraging \cite{Viswanathan:1999optimizing}.

Although the results we reported are limited to walks with 
exponentially distributed waiting times, our formalism can be easily
extended to include the possibility of waiting times with power law
distributions such as observed in
Ref.~\cite{Solomon:1993observation}. Such processes are  
known to allow for sub-diffusive transport regimes
\cite{Metzler:2000PhysRep}. The combination of two power  
law scaling parameters, one for the waiting time and the other for the
duration of flights, indeed yields a richer set of scaling regimes
\cite{Portillo:2011intermittent}, which can be studied within our
framework. 

Our results can on the other hand be readily applied to the regime
$\taur/\taub\ll1$, i.e., such that the waiting times in the scattering state are
typically negligible compared to the ballistic timescale. This is the
regime commonly studied in reference to L\'evy walks.

Our investigation simultaneously opens up new avenues for future
work. Among results to be discussed elsewhere, our formalism can
be used to obtain exact solutions of the mean squared displacement as
a function of time. As discussed already, this is particularly useful
to study transient regimes, such as can be observed when the
distribution of free paths has a cut-off or, more generally, when it
crosses over from one regime to another, e.g. from a power law for
small lengths to exponential decay for large ones, or when the
anomalous regime is masked by normal sub-leading contributions which
may nonetheless dominate over time scales accessible to numerical
computations \cite{Cristadoro:2014measuring}. One can also apply these
ideas to the anomalous photon statistics of blinking quantum dots
\cite{Jung:2002Lineshape, Margolin:2004p14506}. The on/off switchings
of a quantum dot typically exhibit power law distributions. In the limit of 
strong fields, however, the on-times display exponential cutoffs.  

Another interesting regime occurs when, in the positive-recurrent
range of the scaling parameter, $\alpha > 1$, the likelihood of a
transition from a scattering to a propagating state is small, $\epsilon \ll 1$. A
similar perturbative regime arises in the infinite horizon Lorentz gas
in the limit of narrow corridors \cite{Bouchaud:1985p7680}. As is
well-known \cite{Bleher:1992JSP315}, the scaling parameter of the
distribution of free paths has the marginal value $\alpha = 2$, such
that the mean squared displacement asymptotically grows with $t \log
t$. Although it has long been acknowledged that the infinite horizon
Lorentz gas exhibits features 
similar to a L\'evy walk \cite{Levitz:2007gw, barkai:1997p6355}, we
argue that a consistent treatment of this model in such terms is not
possible unless exponentially-distributed waiting times are taken into
account that separate successive jumps. Indeed, the parameter
$\epsilon$, which weights the likelihood of a transition from scattering
to propagating states, is the same parameter that separates the average
relaxation time of the scattering state from the ballistic  timescale,
i.e., $\taub/\taur \propto \epsilon \ll 1$. This is the subject
of a separate publication \cite{Cristadoro:2014Machta}.  

\acknowledgments
We wish to thank Eli Barkai for useful comments and suggestions. This
work was partially supported by FIRB-Project No. RBFR08UH60 
(MIUR, Italy), by SEP-CONACYT Grant No. CB-101246 and DGAPA-UNAM
PAPIIT Grant No. IN117214 (Mexico), and by FRFC convention 2,4592.11
(Belgium). T.G. is financially supported by the (Belgian) FRS-FNRS.

\end{document}